%% file: chbib.tex
\documentclass[USenglish]{article}\usepackage{amsmath,amsthm,amssymb,fullpage,url,xspace,cite,quotes,algorithm2e}
\usepackage{graphicx}
\usepackage[hidelinks]{hyperref}
\newtheorem{theorem}{Theorem}[section]\newtheorem{lemma}[theorem]{Lemma}\newtheorem{corollary}[theorem]{Corollary}
\usepackage{enumitem}
\usepackage{caption}
\usepackage{subcaption}
\title{Cache-Oblivious Parallel Convex Hull in the Binary Forking Model} 

\newcommand{\shih}[1]{}

\begin{document}
\pagenumbering{gobble}
\author{Reilly Browne \thanks{Department of Computer Science,
        Stony Brook University, {\tt rjbrowne@cs.stonybrook.edu}} 
        \and Rezaul Chowdhury \thanks{Department of Computer Science,
        Stony Brook University, {\tt rezaul@cs.stonybrook.edu}}
        \and Shih-Yu Tsai  \thanks{Department of Computer Science,
        Stony Brook University, {\tt shitsai@cs.stonybrook.edu}}
        \and Yimin Zhu \thanks{Department of Computer Science,
        Stony Brook University, {\tt Yimin.Zhu@stonybrook.edu}}}
\maketitle
\begin{abstract}
We present two cache-oblivious sorting-based convex hull algorithms in the Binary Forking Model. The first is an algorithm for a presorted set of points which achieves $O(n)$ work, $O(\log n)$ span, and $O(n/B)$ serial cache complexity, where $B$ is the cache line size. These are all optimal worst-case bounds for cache-oblivious algorithms in the Binary Forking Model.
The second adapts Cole and Ramachandran's cache-oblivious sorting algorithm, matching its properties including achieving  $O(n \log n)$ work, $O(\log n \log \log n)$ span, and $O(n/B \log_M n)$ serial cache complexity. Here $M$ is the size of the private cache.
\end{abstract}

\input{intro}

\input{arxiv/presorted_full}

\input{algorithm_docs/multiwayMerge}

\input{conclusion}

\bibliography{chbib.bib}

\end{document}

%% file: intro.tex
\section{Introduction}
Finding the convex hull of a set of $n$ points in the plane is one of the most fundamental problems in computational geometry. It has wide applications, such as robot motion planning in robotics,  image processing in pattern recognition, and tracking disease epidemics in ethology~\cite{akl1979efficient,dumonteil2013spatial,mercy2016real}.
In the serial setting, there have been several efficient algorithms for constructing convex hulls~\cite{ChanAlgorithm,farach2018streaming,Graham1972AnEA,Preparata1977}, both in 2-dimensions and in the more general problem of $d$-dimensions. In the parallel setting, a bunch of efficient convex hull algorithms have been developed  for a set of $d$-dimensional points~\cite{amato1994parallel,atallah1989cascading,ATALLAH1986,blelloch2020randomized,gupta2003faster,miller1988efficient,reif1992optimal}. 
There are also many parallel convex hull algorithms specifically for presorted or unsorted points in 2-dimensions~\cite{berkman1996fast,chen1995efficient,ghouse1991place,miller1988efficient}. 

In this paper, we focus on designing parallel algorithms for the convex hull of a set of points in 2-dimensions in the \textbf{\textit{binary-forking model}} \cite{BlellochFiGuSu2019}. In particular, we present \textbf{\textit{cache-oblivious}} \cite{Frigo} \textbf{\textit{parallel}} algorithms for both presorted (by $x$-coordinate) and unsorted points with optimal worst-case \textbf{\textit{work}} (see next paragraph) with respect to the size of the input. We achieve optimal \textbf{\textit{span}} (see next paragraph) in the presorted case and match the best-known span for sorting in the unsorted case for the binary-forking model.
%

We use the \textit{work-span} model \cite{CormenLeRiSt2009} to analyze the performance of parallel algorithms. The \textit{work}, $W(n)$ of a parallel algorithm is defined as the total number of CPU operations it performs when it is executed on a single processor. Its \textit{span}, $S(n)$ on the other hand, is the maximum number of operations performed by any single processor when the program runs on an unlimited number of processors.


The \textit{binary-forking} model~\cite{acar2000data, ben2016parallel, blelloch2008provably, blelloch2011scheduling,BlellochFiGuSu2019} realistically captures the performance of parallel algorithms designed for modern multicore shared-memory machines. Its formal definition is in~\cite{BlellochFiGuSu2019}. In this model, the computation starts with a single thread, and then threads are dynamically and asynchronously created by some existing threads as time progresses. The creation of threads is based on the spawn/fork action: a thread can spawn/fork a concurrent asynchronous child thread while it continues its task. Note that the forks can happen recursively. The model also includes a join operation to synchronize the threads and an atomic test-and-set (TS) instruction. This model can be viewed as an extension of the binary fork-join model~\cite{CormenLeRiSt2009} which does not include the TS instruction.
This model and its variants~\cite{ acar2000data, agrawal2014provably,blelloch2008provably, blelloch2004effectively,BlumodeLeiserson,CormenLeRiSt2009} are widely used in many parallel programming languages or environments such as Intel TBB~\cite{tbb}, the Microsoft Task Parallel Library~\cite{micpparlib}, Cilk~\cite{frigo1998implementation}, and the Java fork-join framework~\cite{forkjoin}. 

 The binary-forking model is an ideal candidate for modeling parallel computations on modern architectures when compared with the closely related PRAM model~\cite{jaja1997introduction}. The main difference between the binary-forking model and the PRAM model is synchronicity. The binary-forking model allows asynchronous thread creation; in the PRAM model, all processors work in synchronous steps. PRAM does not model modern architectures well because they utilize new techniques such as multiple caches, branch prediction, and many more, which lead to many asynchronous events such as varying clock speed, cache misses, etc~\cite{BlellochFiGuSu2019}. 
 Any algorithm designed for the PRAM model can be transformed into an algorithm for the binary-forking model at the cost of an $O(\log n)$-factor blow-up in the span while keeping the work asymptotically the same as in the PRAM model.

We also employ the use of the \textit{cache-oblivious} model, first described by Frigo et al. \cite{Frigo}. In this model, memory is assumed to have two layers: a cache of size $M$ and a main memory of unlimited size. The memory is split into blocks of size $B$, and every time the processor tries to access a data point that is not in the cache it incurs a cache miss and the block containing the data point is copied into the cache from the main memory. When copying into a cache that is already full,  an old block is evicted to make space for the new block. However, in contrast to the cache-aware model, cache-oblivious algorithms do not use the knowledge of the values of $M$ and $B$. In both models, the cache complexity of an algorithm is measured in terms of the number of cache misses it incurs and is referred to as cache-optimal if it incurs the fewest possible cache misses asymptotically.

There are several ways to expand the cache-oblivious model to parallel computation. The performance of our algorithms are analyzed for private caches, using the same model as in Cole and Ramachandran \cite{ColeRamachandran}. It is similar to the parallel external memory (PEM) model of Arge et al. \cite{ArgeGoodrich}, with the primary difference being in the type of parallelism used. The PEM model uses bulk-synchrony for synchronization whereas we use binary forking. However, the use of private caches is almost identical. Each processor has a private cache of size $M$ which consists of blocks of size $B$ and the main memory of arbitrary size is shared amongst all processors. Several problems have been studied in the PEM model, particularly geometric problems including convex hull \cite{Sitchinava}. Sitchinava \cite{Sitchinava} showed that Atallah and Goodrich's \cite{AtallahGoodrichCache} $O(n \log n)$ work and $O(\log n)$ span algorithm on presorted points  matches the cache misses of sorting for the PEM model. This implies a convex hull algorithm for unsorted points that matches sorting bounds. However, in the binary-forking model and the binary fork-join model, the span of their algorithm becomes $O(\log ^2 n)$ which is dominated by thread-spawning at each level in the $O(\log n)$ height recursion. We improve upon this result by presenting an algorithm with $O(\log n)$ span and $O(n)$ work for presorted points while preserving cache-obliviousness. 

In terms of cache analysis for convex hulls, one of the earliest serial algorithms developed by Graham \cite{Graham1972AnEA}, when combined with a cache-oblivious sorting algorithm, could achieve $O(n/B \log _M n)$ cache misses. Arge and Miltersen \cite{ArgeMiltersen} showed that this bound is optimal for non-output sensitive convex hull algorithms in the cache-aware model, which carries over to the cache-oblivious model. When output sensitivity is accounted for, where $h$ is the number of points comprising the convex hull, this bound decreases to $O(n/B \log _{M/B} (h/B))$, as is achieved by Goodrich et al. \cite{GoodrichCache} for the external memory model. In terms of parallel cache-oblivious algorithms, Sharma and Sen \cite{SharmaSen} presented a randomized CRCW algorithm that achieves expected $O(n/B \log _M n)$ cache misses and expected $O(\log n \log \log n)$ span.




%

\subparagraph*{[Our Contributions.]} In summary, we have the following results:
\begin{itemize}
    \item For finding the convex of a set of presorted points (by x-coordinate), we give a deterministic cache-oblivious algorithm that uses $n^{1/5}$-way divide and conquer, performing $O(n)$ work in $O(\log n)$ span and incurring $O(n/B)$ cache misses. It is optimal across work, span and cache complexity for the problem of finding the convex hull of presorted points in the binary-forking model.

    \item For finding the convex hull of unsorted points, we give a  deterministic cache-oblivious algorithm based on multi-way merging which performs worst-case optimal $O(n \log n)$ work in $O( \log n \log\log n)$ span, and achieves  optimal parallel cache complexity. To the best of our knowledge, this is the first deterministic cache-efficient 2D convex hull algorithm for unsorted points in the binary-forking model with the lowest span. Its span outperforms the recently proposed randomized incremental convex hull algorithm for the binary-forking model as well \cite{blelloch2020randomized}. It matches the best known bounds for span, work, and cache-efficiency for sorting, which means it matches the previous result but simply requires changing a sorting algorithm instead of implementing two complex procedures.
    
    \item All our bounds hold for the binary fork-join model as well since neither algorithm uses the atomic test-and-set operation which is unique to the binary-forking model.

\end{itemize}



%% file: arxiv/presorted_full.tex
\section{A Cache-Optimal \texorpdfstring{$O(\log n)$}{O(log n)} Span Algorithm for Presorted Points}

We present a cache-oblivious divide-and-conquer algorithm for finding the upper convex hull of an array of presorted points. It achieves $O(\log n)$ span with $O(n)$ work and $O(n/B)$ cache misses, which are optimal bounds for work, span, and cache efficiency for the convex hull problem in the binary-forking model. As is common practice, we describe a procedure which finds the "upper hull", the set of extreme points which lie above the line from the leftmost and rightmost points. Repeating this procedure on the point set rotated by 180 degrees yields the entire convex hull. We consider our algorithm to be one of several to build off of the recursive structure of Atallah and Goodrich's \cite{ATALLAH1986} $O(n \log n)$ work algorithm. Atallah and Goodrich's algorithm is a $\sqrt{n}$-way divide-and-conquer algorithm. To perform the merges, it uses Overmars and Van Leuwen's \cite{OVERMARS1981166} technique for finding the upper common tangent of two disjoint convex polygons to compare the $\sqrt{n}$ subproblem solutions and determine which points from each should remain. At each layer in this recursion, $\binom{\sqrt{n}}{2} = O(n)$ applications of this technique are used at the cost of $O(\log n)$, giving the recurrence $T(n) = \sqrt{n} T(\sqrt{n}) + O(n \log n)$ for the total work. The span of this method is $O(\log n)$, even under the more restrictive binary-forking model where $O(\log n)$ is optimal for any problem requiring $\Omega(n)$ work, as spawning $\Omega(n)$ threads incurs $\Omega(\log n)$ span. However, here the work remains suboptimal.

Goodrich \cite{goodrich1987finding} presented a method to obtain $O(n)$ work using a tree data structure which they refer to as a ``hull tree''. The hull tree data structure stores the convex hull of a set of points in a binary tree that permits a simulation of the Overmars and Van Leuwen's procedure and an efficient means of ``trimming off'' points that have been eliminated and unifying subproblem solutions under a new tree. The issue is that using this data structure incurs many cache misses because the tree cannot guarantee any data locality. Chen \cite{chen1995efficient} also uses a tree-based approach to achieve $O(n)$ work, which encounters the same issue.

The first step to ensure data locality while reducing to $O(n)$ work is to decrease the number of subproblems so that the pairwise hull comparisons do not take $O(n \log n)$ work. In fact, we want this work from these comparisons to total $O(n/B)$. This can be done by choosing a different exponent. We will use $n^{1/6}$-way divide-and-conquer, which means that comparing the subproblem solutions will take $O(n^{2/6} \log n)$ work, which allows us to use the tall cache assumption ($M = \Omega(B^2)$) to bound the total work from these comparisons by $O(n/B)$. Once the subproblems are compared, we can eliminate all of the points which have been determined to be within the hull by shifting all of the remaining points to the front of the array (using a prefix sum). This can be done cache-obliviously \cite{blelloch2010low}, incurring $O(n/B)$ cache misses for each recursion layer. However, this comes out to be $O(n \log \log n)$ work total. So, if we are to maintain data locality, we have to determine a way to eliminate points without having to shift everything down and incur too much work.

Instead of deleting points immediately, Algorithm \ref{alg:presorted}, (pseudocode in appendix) uses an array of size $O(\frac{n}{\log n})$ which stores a set of x-coordinate intervals corresponding to the points which must be deleted. We store these intervals as a list of ``starts'' and ``ends'' which correspond to the beginning and ending of each deletion interval all sorted by x-coordinate. For an interval $I$, consider the largest interval $I'$ for which $I \subseteq I'$. As an invariant, we will store in $start(I)$ a pointer to $start(I')$ and in $end(I)$ a pointer to $end(I')$. This will crucially allow us to get from a point that is within one of these intervals to the nearest point which is not within any of these deletion intervals in $O(1)$ time, which lets us simulate the Overmars and Van Leuwen \cite{OVERMARS1981166} technique on subproblem solutions. By primarily operating on this array of intervals, we can avoid incurring $O(n)$ work at every layer and instead amortize to $O(n)$ total work across all layers.

\begin{algorithm}
\small
\caption{Cache-Optimal algorithm for Presorted Points}\label{alg:presorted}
\KwData{$A$, a subset of points that are sorted by their $x$-coordinate. $n$, the number of points in $A$. $N$, the number of points in the original input, which is a superset of $A$. $D_{alloted}$, an allotted section of global array $D$. $n_{alloted}$, the size of alloted section $D_{alloted}$}
\KwResult{The upper hull of $A$}
    \textbf{if $n < \log N$ then} solve using a serial algorithm (such as Graham scan) \;
    \Else{
  Split $A$ into $n^{1/6}$ subproblems of size roughly $n^{5/6}$. Allot a contiguous segment of $\frac{n_{alloted} - 2n^{1/6}}{n^{1/6}}$ slots of $D_{alloted}$ to each subproblem (in x-order). Solve these recursively\; 
    Find the common tangents between each pair of subproblems  using the modified Overmars and Van Leuwen technique (See Theorem \ref{thm:overmars}). Store in a local array of tangents, $\mathcal{T}$ \;
       Comparing the elements of $\mathcal{T}$ for each subproblem, determine the local set of new dividers, $D_{new}$. Additionally store the dividers in the $2n^{1/6}$ slots at the end of $D_{alloted}$ and another copy in $D_{new}$\;
    Merge the list of old and new dividers by x-order in $D_{alloted}$\;
    For all $d \in D_{new}$, find d's copy in $D_{alloted}$. Store in $d.\textrm{\sc{\em{Rank}}}$ the copy's rank in $D_{alloted}$\;
    For each (contiguous) left-right pair $(d_l, d_r) \subset D_{new}$, mark the dividers $D_{alloted}[d_l.\textrm{\sc{\em{Rank}}},d_r.\textrm{\sc{\em{Rank}}}]$ by setting their \textrm{\sc{\em{Wrapper}}} to $D_{alloted}[d_l.\textrm{\sc{\em{Rank}}}]$ if it is a \textrm{\sc{\em{Left}}} divider, $D_{alloted}[d_r.\textrm{\sc{\em{Rank}}}]$ if \textrm{\sc{\em{Right}}} \;
    For each element $d$ in $D_{new}$, mark the points in $A$ between its $D_{alloted}$ copy's \textrm{\sc{\em{Landing}}} flag and the $D_{alloted}$ copy's neighbor ($D_{alloted}[d.\textrm{\sc{\em{Rank}}}+1]$ for \textrm{\sc{\em{Left}}}, $D_{alloted}[d.\textrm{\sc{\em{Rank}}}-1]$ for \textrm{\sc{\em{Right}}})\;
    Set the \textrm{\sc{\em{Rank}}} flag of all the copies of $D_{new}$ dividers to 1 if  \textrm{\sc{\em{Left}}} and -1 if  \textrm{\sc{\em{Right}}}. Then perform a prefix sum on the  \textrm{\sc{\em{Rank}}} flags, this determines if a divider is a subset of another interval.
    For each element $d$ in $D_{alloted}$ that is a  \textrm{\sc{\em{Right}}} divider with  $d.\textrm{\sc{\em{Rank}}}$ greater than 0 and $d.\textrm{\sc{\em{Marked}}}$ set to false, mark the points in the adjacent interval for deletion and set  $d.\textrm{\sc{\em{Rank}}}$ to true. 
     
    \textbf{if $n=N$ then} Use prefix sums to determine the ranks of the unmarked (extreme) points and delete the marked points\;
    }
    
\end{algorithm}

\subsection{Data Structure Invariants}

We use three data types in our arrays: \textbf{points}, \textbf{dividers}, and \textbf{tangents}. Each type is constant-sized, so the arrays have space linear in their respective numbers of elements.

\textbf{Points} are based on the input, with some extra attributes assigned to each point so that if it is not an extreme point, it can be marked instead of immediately removed. In addition to $x$- and $y$-coordinates, the point contains a \textrm{\sc{\em{Rank}}} integer flag which is initialized to 1 and two pointers to dividers called $\textrm{\sc{\em{LeftPar}}}$ and \textrm{\sc{\em{RightPar}}}. $\textrm{\sc{\em{LeftPar}}}$ and \textrm{\sc{\em{RightPar}}} are initialized to \textrm{\sc{\em{Null}}}, but when they are set, they indicate that the point is within some deletion interval $I$. $\textrm{\sc{\em{LeftPar}}}$ points to a divider corresponding to the start of $I$, whereas \textrm{\sc{\em{RightPar}}} points to a divider corresponding to the end of $I$.

\textbf{Dividers} represent the \textit{boundaries} of deletion intervals. If the divider represents the beginning of some interval $I$, we will refer to it as $L_I$. If the divider refers to the end of $I$, then we will refer to it as $R_I$. The divider contains an $x$-coordinate, a pointer \textrm{\sc{\em{Landing}}} that points to the point element  corresponding to the start/end of the interval (but not within the interval), a pointer \textrm{\sc{\em{Wrapper}}} to a divider from an interval $I'$ that is a superset of $I$ ($L_I'$ for $L_I$ and $R_I'$ for $R_I$), an integer flag \textrm{\sc{\em{Rank}}} for use in prefix sums, a bit flag  \textrm{\sc{\em{Marked}}} to denote whether  \textrm{\sc{\em{Landing}}} has been marked, and a bit flag indicating whether the divider is a ``\textrm{\sc{\em{Left}}}'' divider (the start of an interval) or a ``\textrm{\sc{\em{Right}}}'' divider (the end of an interval). All of these are initialized to 0 or \textrm{\sc{\em{Null}}}.

\textbf{Tangents} represent an upper common tangent between the upper hulls of two subsections. An upper common tangent between two hulls is the line that lies above all points in each hull and which intersects the boundaries of each hull in exactly one contiguous section. This is typically expressed as a pair of extreme points, one from each hull. These are used to determine which points would be deleted if the two hulls were merged together. In our representation, each tangent simply contains 2 pointers to points, one for the point of tangency of the leftward of the two hulls, and one for the point of tangency of the rightward of the two hulls. Both pointers are set to \textrm{\sc{\em{Null}}} to start.

Our presorted algorithm will utilize 4 arrays of these types: 2 permanent arrays (of points, $A$, and of dividers $D$) and 2 temporary arrays specific to the particular recursions (of tangents, $\mathcal{T}$, and of dividers, $D_{new}$).

There are 5 invariants that are maintained with respect to $D$ that will allow us to bound our work by $O(n)$ and our cache misses by $O(n/B)$.

\begin{enumerate}[align=left,label=\textbf{Invariant \arabic*},ref=\arabic*]\setlength{\itemindent}{.35cm}
    \item \label{invariant:1} The size of $D$ must be $O(\frac{n}{\log n})$.
    \item \label{invariant:2} For any interval $I$ with dividers $L_I, R_I \in D$, and the largest subset $I'$ such that $I \subseteq I'$, $L_{I}.\textrm{\sc{\em{Wrapper}}}= L_{I'}$ and $R_{I}.\textrm{\sc{\em{Wrapper}}}= R_{I'}$. In the case of ties, take $I'$ to be that was most recently added.
    \item \label{invariant:3} For any two intervals $I_1,I_2$ with dividers in $D$, either $I_1 \cap I_2 = \emptyset$, $I_1 \subseteq I_2$, or $I_2 \subseteq I_1$.
    \item \label{invariant:4} For any point $p$ which is in at least one interval, for one such interval $I$, $p.\textrm{\sc{\em{LeftPar}}} = L_I$ and $p.\textrm{\sc{\em{RightPar}}} = R_I$. 
    \item \label{invariant:5} For any interval $I$ for which $L_I.\textrm{\sc{\em{Wrapper}}} = L_I$ and $R_I.\textrm{\sc{\em{Wrapper}}} = R_I$, then neither $L_I.\textrm{\sc{\em{Landing}}}$ nor $R_I.\textrm{\sc{\em{Landing}}}$ are elements of any interval.
\end{enumerate}

\begin{figure}[t]
    \centering
    \begin{subfigure}[b]{0.3\textwidth}
        \centering
        \includegraphics[height=.5\textwidth]{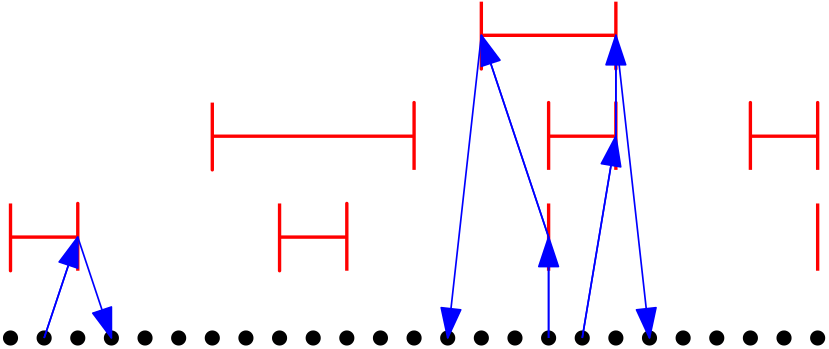}
        \caption{}
        \label{fig:intervaltraversal}
    \end{subfigure}
    \begin{subfigure}[b]{0.65\textwidth}
        \centering
        \includegraphics[height=.5\textwidth]{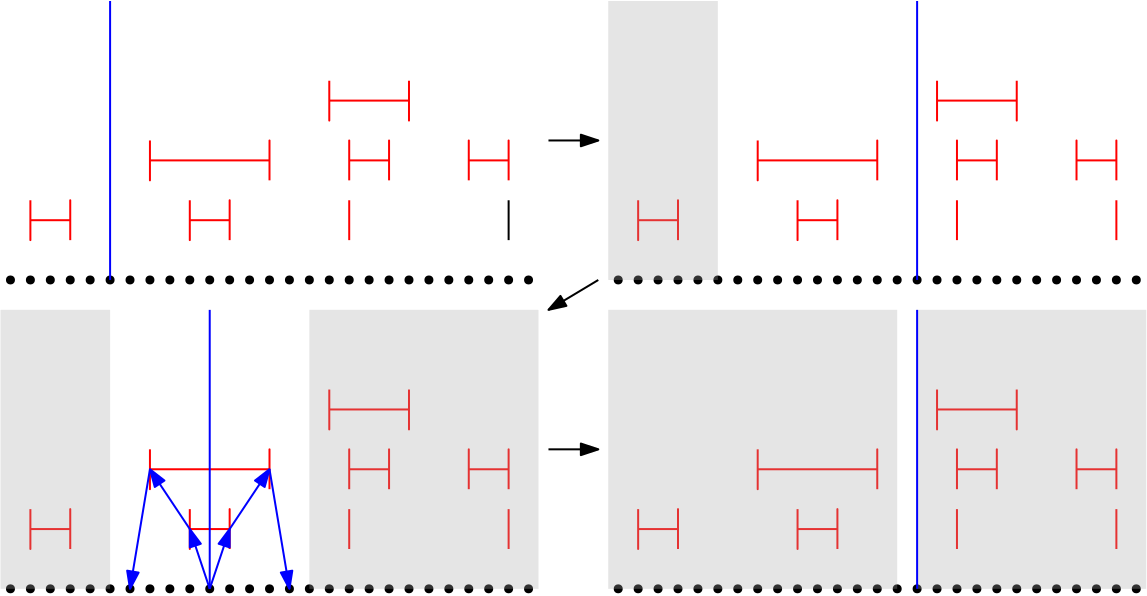}
        \caption{}
        \label{fig:overmarssim}
    \end{subfigure}
    \caption{(a) Example of paths from marked points to points that have not been marked; (b) Simulating the Overmars and Van Leuwen \cite{OVERMARS1981166} procedure using the divider array. Only one of the two hulls are depicted. The blue line depicts the current point of consideration and the green the previous point of consideration.}
    \label{fig:my_label}
\end{figure}

We give an example depiction of this structure in Figure~\ref{fig:intervaltraversal}. We show this in the context of paths from points that are marked for deletion (lie in some interval) to a nearest point to the left or right which is not in that interval. Assuming the invariants hold, we can claim the following theorem.

\begin{theorem}\label{thm:overmars}
    Given two convex hulls $C_1$, $C_2$, that are both expressed by an array of points $A$ and an array of dividers $D$, and that are separable by a vertical line, the upper common tangent of $C_1$ and $C_2$ can be computed in $O(\log n)$ time serially.
\end{theorem}

\begin{proof}
We assume that $A$ and $D$ abide by the 5 invariants listed above. We will simulate Overmars and Van Leuwen \cite{OVERMARS1981166} on $A$, making use of $D$ in the case that we accidentally land on a point which has been marked for deletion i.e. is not in either $C_1$ or $C_2$. Since we can proceed as usual if there are no dividers (and hence no intervals of deletion), we will consider only the case in which there have been points marked for deletion.

For any unmarked point, we will consider the \textit{left path} and the \textit{right path} in the divider structure. These paths follow down pointers from the source, so we will repreent the points in the path using $\to$ as a delimiter, indicating we are going to the point pointed to by that flag. These paths land on the nearest unmarked point to the left and right respectively. The left path is either 
$p\to LEFTPAR \to LANDING$ or, if $p.LEFTPAR$ points to the $L_I$ for an interval $I$ which is the subset of a larger interval $I'$, then the path is $p\to LEFTPAR \to WRAPPER \to LANDING$. By invariant \ref{invariant:5}, we know that this path will land on a point which has not been marked. Specifically, this point is the point which is immediately outside $I$'s (or $I'$'s) interval on the left side. Since $p \in I$ ($p \in I')$, any point between the end of the left path and $p$ must be marked for deletion. Therefore, the left path gives the nearest point to the left. By symmetry, the same applies for the right path, given as  $p \to RIGHTPAR \to LANDING$ or $p \to RIGHTPAR \to WRAPPER \to LANDING$. These paths are shown in blue in Figure \ref{fig:intervaltraversal}.


Using the left and right paths, we can deal with the case of points marked for deletion. The Overmars and Van Leuwen \cite{OVERMARS1981166} technique relies on the ability to eliminate half of the remaining candidate points (for the line of tangency, not the hull) from either $C_1$, $C_2$ or both at every step, which we also guarantee. This gives us the same $O(\log n)$ time guarantee since determining the points to eliminate from consideration for the upper common tangent takes $O(1)$ time.

In the first part of the Overmars and Van Leuwen \cite{OVERMARS1981166} technique, the point with maximum y-coordinate, $p_{top}$, is found for each hull using a binary search. Consider $C_1.$ Each step in this search determines whether $p_{top}$ is to the left or right of the current point, $p$, by comparing it to its predecessor and successor in $C_1$. If the successor to $p$ in the array $A$ is marked for deletion, take the right path from this to get the true successor of $p$. For the predecessor, take the left path. If $p$ itself is marked for deletion, then consider both the point at the end of the left path and the point at the end of the right path. These comparisons will either lead to the candidate set contracting to either the left or right of both or we will know that the top is one of the two, since the top point cannot be in the deleted interval. In either case, at least half of the points are eliminated from consideration since $p$ lies at the midpoint of the candidates. We show an example search in Figure~\ref{fig:overmarssim}. 

The second part of the technique is another binary search, but this one relies on both hulls. At every step, we consider the midpoint of the remaining candidates in $C_1$ (and also for $C_2)$. Just as before, if this point is marked, we instead consider the points at the end of the left path and the right path. The details of how to compare two points that are being considered (one from $C_1$ and one from $C_2$) are given in \cite{OVERMARS1981166}. When the comparison is done, we can eliminate all candidates to the left of or to the right of at least one of the considered points. This holds for our case even if we have to consider 2 points from $C_1$ and two points from $C_2$. Let $l_1,l_2$ be the points from following the left paths and $r_1,r_2$ be the points from following the right paths. We represent the comparisons as  pairs $(l_1,l_2),(l_1,r_2),(r_1,l_2),(r_1,r_2)$. Each comparison specifies that for at least one of the points, all points in its respective hull to either the left or the right cannot be a point of tangency. Whichever way the comparisons lie, one of the two hulls $C_i$ will have both $l_i$ and $r_i$ able to eliminate points. If this is the case, then the argument from the search for $p_{top}$ applies here. If only one of $C_1$ or $C_2$ has 2 points to consider from landing in an interval, without loss of generality consider $C_1$ to have the 2 points of consideration. Either both points in $C_1$ can eliminate candidates in the search, which gives us the same as the previous case or the point from $C_2$ can eliminate half to its left or right. In either case, half of the remaining candidates from  $C_1$ or $C_2$ have been eliminated. If both $C_1$ and $C_2$ have 1 point to consider, it follows directly from the original Overmars and Van Leuwen technique.

Therefore, we can guarantee that only $O(\log n)$ comparisons need to be made to find the upper common tangent between $C_1$ and $C_2$. \end{proof}

\subsection{Maintaining the invariants}

Now that we have a method to compare the hulls assuming the invariants hold, we will show how to merge the hulls, keeping in mind that we also must maintain the invariants. Lemma \ref{lma:dividercreation} will show how to maintain invariants \ref{invariant:1} and \ref{invariant:3}, while Lemma \ref{lma:markingpoints} will show how to maintain invariants \ref{invariant:2}, \ref{invariant:4}, and \ref{invariant:5}.

\begin{lemma} \label{lma:dividercreation}
The set of new dividers $D_{new}$ for a given recursion layer (of size $n$) can be determined in $O(\log n)$ span with $O(n^{2/6} \log n)$ work (and cache misses)
\end{lemma}

\begin{proof}
We will first calculate the $O(n^{2/6})$ common tangents, storing them in a 2D array $\mathcal{T}$. We use the smallest and largest sloped tangents (as in Atallah and Goodrich \cite{ATALLAH1986}) to determine which points should be removed from each subproblem hull, or subhull. For each subhull, there will be some section starting with the leftmost point and some section starting with the rightmost point which will need to be removed, leaving some contiguous set of points in the center, or no points from this hull. If there is such an set, we use a \textrm{\sc{\em{Right}}} divider with \textrm{\sc{\em{Landing}}} set to the first (leftmost) unremovable point and a \textrm{\sc{\em{Left}}} divider with \textrm{\sc{\em{Landing}}} set to the last (rightmost) unremovable point. If there are no unremovable points from this subhull, then do not add any divider. We can omit dividers for the leftmost and rightmost points among all of these subhulls, since they will not correspond to intervals of deletion (they are extreme points). We can store these all in an array of size $2n^{1/6}-2$, with gaps allowed for nonexisting dividers. Set the \textrm{\sc{\em{Wrapper}}} for each of these dividers to itself.

At this point, $D_{new} \cup D$ has satisfied invariants \ref{invariant:1} and \ref{invariant:3}. \ref{invariant:1} is satified since $D$ must be of size at most $O(\frac{n}{5/6 \log n})$ and the addition of $2n^{1/6}-2$ dividers does not affect this asymptotically. \ref{invariant:3} is satisfied by induction, since when $D$ is empty, this holds since it holds for $D_{new}$ alone, and assuming it holds for $D \neq \emptyset$, since $D_{new}$ does not have dividers that "land" inside previously marked points, they cannot intersect with any previous intervals corresponding to $D$ without entirely containing such an interval.
\end{proof}

\begin{lemma}\label{lma:dividerintegration}
The set of new dividers $D_{new}$ for a given recursion layer (of size $n$) can be integrated into the full set of dividers $D$ in $O(\log n)$ span with $O(\frac{n}{\log n})$ work and $O(\frac{n}{B \log n})$ cache misses.
\end{lemma}
\begin{proof}
We can integrate these using a merge routine. Specifically, we will be merging a copy of $D_{new}$ into $D_{alloted}$, a set of contiguous slots in $D$ that have been allotted to this subproblem. After alloting for each of the subproblems, there are $2n^{1/6}$ slots remaining at the end, allowing us to copy $D_{new}$ there. Before doing so, we will compress $D_{new}$ so they occupy contiguous positions. Only then will we copy them over to the end of $D_{alloted}$. These both can be done using prefix sums \cite{blelloch2010low}, which takes $O(n^{1/6})$ work and $O(n^{1/6}/B)$ cache misses. These calculations will use the \textrm{\sc{\em{Rank}}} flag. \end{proof}

\begin{lemma} \label{lma:markingpoints}
All non-extreme points can be marked for deletion with $O(n)$ work and $O(n/B)$ cache misses, amortized.
\end{lemma}

\begin{proof}

We show that every cache miss incurred by the process of marking points for deletion can be charged either to the binary searches (which we show to be $O(n/B)$ in Theorem \ref{thm:presorted_final}) or to the total number of blocks representing the points. We also show that the remaining invariants hold. 


Now that the array $D_{new}$ has copies of its elements integrated into $D_{alloted}$, we can start by satisfying invariant \ref{invariant:2}. For each adjacent pair in $D_{new}$ corresponding to an interval of deletion $I$, conduct binary searches on $D_{alloted}$ for their copies. Between these two copies will be a set of dividers corresponding to intervals which are subsets of $I$. We have already marked the points inside these intervals, so we just need to mark what is left. We start with the points between the beginning of $I$ and its first subset, and the points between the last subset and the end of $I$. To determine these subsets, we will store the rank in $D_{alloted}$ of the copies in the \textrm{\sc{\em{Rank}}} flags of the $D_{new}$ elements. For the ``in-between'' dividers in $D_{alloted}$ which are \textrm{\sc{\em{Left}}}, set their \textrm{\sc{\em{Wrapper}}} pointers to the \textrm{\sc{\em{Left}}} element of the $D_{new}$ pair, specifically the copy in $D_{alloted}$. For the \textrm{\sc{\em{Right}}} divider, do the same with the \textrm{\sc{\em{Right}}} copy. This will satisfy invariant \ref{invariant:2} by the same principle of induction as before.  All that remains is to mark the points inbetween subintervals of $I$. These points are distinguished by being between dividers which have their  \textrm{\sc{\em{Marked}}} flag set to false (the default) and which are themselves between a \textrm{\sc{\em{Left}}}-\textrm{\sc{\em{Right}}}divider pair from $D_{new}$. We will set each point's \textrm{\sc{\em{LeftPar}}} to the left divider of the subinterval after the point and the \textrm{\sc{\em{RightPar}}} to the right divider of the subinterval before the point. This will still let the point reach to the nearest unmarked point, since the subintervals all point to the largest superset.

The marking procedure is cache efficient, incurring $O(\frac{n}{B \log n})$ cache misses at each recursion layer. There can be additional cache misses incurred due to the deletion marking intervals not aligning with blocks, but these misalignments can only happen at the edges of an interval, which means we can charge 2 cache misses to one of the binary searches used to compute that divider earlier. There may not be a cache miss corresponding to the binary search if the divider came from a subproblem of size less than $M$. If this is the case, we can argue that if two different dividers corresponding the points within the same block have this occur, then unless they are subsumed by the same larger interval later, then whichever intervals subsume them must not take the entire block, which means one of their endpoints lands in the block and we can amortize the accesses to the binary searches for subsuming intervals. The searches are not cache efficient, but only incur $O(n^{1/6} \log n)$ work, which is dominated by the binary searches used for comparing the subhulls, which are analyzed in Theorem~\ref{thm:presorted_final}.

At this point, invariant \ref{invariant:5} has already been satisfied. Now that the \textrm{\sc{\em{Wrapper}}} flags have been set for intervals that are subsets, we can note that from before the \textrm{\sc{\em{Landing}}} flags for $D_{new}$ elements will not land inside previously marked points. This means that the only way a divider $d$ could have a \textrm{\sc{\em{Landing}}} which has been marked for deletion is if that landing were in a new interval, in which case the interval of $d$, $I$, is a subset of the interval that marked $d.\textrm{\sc{\em{Landing}}}$, by \ref{invariant:3}.

All that remains is invariant \ref{invariant:4}. To maintain this, we mark the points for deletion. To ensure that we do not mark twice, we only consider intervals that were previously unmarked in the previous iteration, or those that are between dividers from $D_{new}$.  Consider a divider $d\in D_{new}$ Let $d'$ be the corresponding divider to $d$ that together represent interval $I$. Looking at the copy of $d$  in $D_{alloted}$, we take the neighboring divider: $D_{alloted}[d.\textrm{\sc{\em{Rank}}}+1]$ if $d$ is a \textrm{\sc{\em{Left}}} divider and $D_{alloted}[d.\textrm{\sc{\em{Rank}}}-1]$ if $d$ is a \textrm{\sc{\em{Right}}} divider. These dividers have landing spots, between which are points will be marked for deletion (exclude the landing spot of a $D_{new}$). Consider the $d$ to be \textrm{\sc{\em{Left}}}. Set $\textrm{\sc{\em{LeftPar}}}$ to $D_{alloted}[d.\textrm{\sc{\em{Rank}}}]$ and \textrm{\sc{\em{RightPar}}} to $D_{alloted}[d'.\textrm{\sc{\em{Rank}}}]$. Do the reverse if $d$ is \textrm{\sc{\em{Right}}}. This will satisfy invariant \ref{invariant:4}. We will mark the points in parallel, in groups of size at most $\log n$.

\begin{figure}
    \centering
    \includegraphics[width=0.5\linewidth]{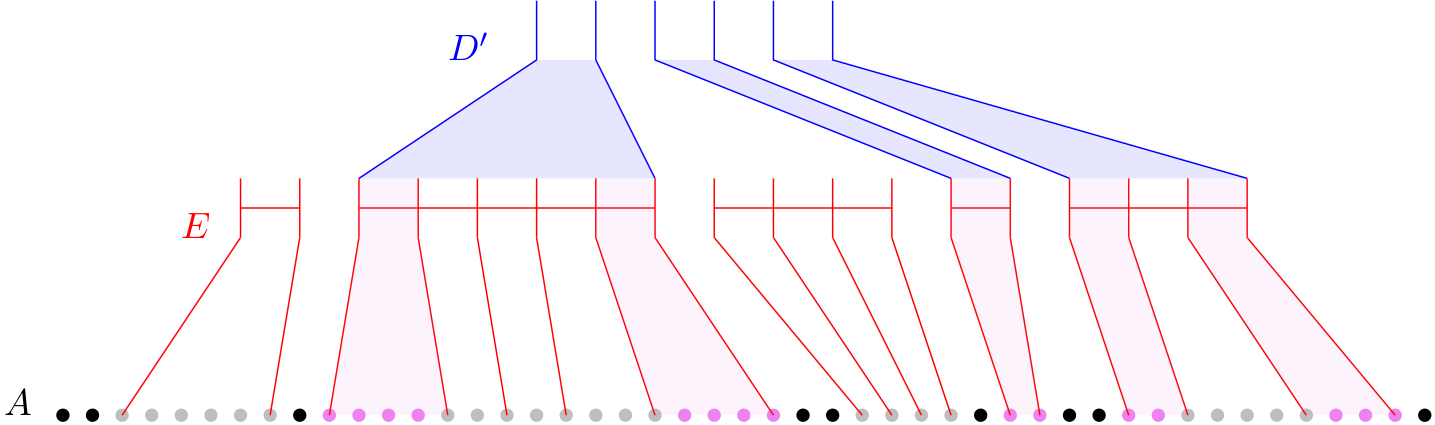}
    \label{fig:deletion_marking}
    \caption{Each pair in $D_{new}$ binary searches for its copy, marks its subsidiaries and then marks the ``outermost layer'' of points.}
\end{figure}

Since only points in "new" intervals are marked, this means that each point is marked at most 1 time, allowing us to amortize the work of marking to $O(n)$. We address some issues due to data alignment in \ref{sec:cachepresort}, but note here that the marking of points is done to adjacent points. We can charge 2 cache misses from each of these to the elements of $D_{new}$, which allows us ignore the issues of alignment on the ends of these intervals. This allows us to consider the "middles" of these intervals, which are access cache-efficiently. These will amortize to $O(n/B)$.\end{proof}

\subsection{Work, Span, and Cache Analysis}

\begin{theorem}\label{thm:presorted_final}
    Algorithm \ref{alg:presorted} achieves $O(\log n)$ span with $O(n)$ work and incurring $O(n/B)$ cache misses
\end{theorem}

\begin{proof}
First, we show span. To start, Algorithm \ref{alg:presorted} finds the convex hull of $O(\frac{n}{\log n})$ subhulls at the lowest level in the recursion layer. Each individual subhull is done serially using a serial algorithm. The Graham scan\cite{Graham1972AnEA} is ideal since it achieves $O(n/B)$ cache misses for presorted input. With each subhull having size at most $\log n$, this takes $O(\log n)$ span. From here, merges are used. The merges start with finding the new dividers $D_{new}$, which takes $O(\log n)$ span from Lemma \ref{lma:dividercreation}. From there, integrating them into the full set of dividers takes $O(\log n)$ span from Lemma \ref{lma:dividerintegration}. Finally, marking the points for deletion takes $O(\log n)$ span due to the binary searches. Marking the points and old dividers takes no more than $O(\log n)$ span since each point/divider can be done in parallel. This gives us the following recursion for subproblems of size $\tilde{n}$, for some positive constant $c_0$:

$$T_\infty(\tilde{n}) \leq T_\infty(\tilde{n}^{1/6}) + c_0 \log \tilde{n} $$

For $\tilde{n} = n$, this expands to $c_0 \log n + \frac{c_0 \log n}{6} + \frac{c_0 \log n}{6^2} + . . . $ which is clearly upper-bounded by $2 c_0 \log n = O(\log n)$ The final step, using a prefix sum to align the unmarked points into adjacent array positions, takes only an additional $O(\log n)$ span \cite{blelloch2010low}.

For the work, the total work is dominated by the final prefix sum, $O(n)$ \cite{blelloch2010low}. The $O(\frac{n}{\log n})$ base cases of size $\log n$ cost $O(n)$ work in total when Graham's algorithm is used. From Lemma \ref{lma:markingpoints}, we know that marking points for deletion costs only $O(n)$ work amortized. From here, it suffices to show that all other operations cost $O(\frac{n}{\log n})$ work per recursion layer. The work from Lemmas \ref{lma:dividercreation} and \ref{lma:dividerintegration} are dominated by $O(\frac{n}{\log n})$, which leaves the work on the divider array leading up to point marking. Setting the old dividers' \textrm{\sc{\em{Wrapper}}} flags is linear in the size of $D_{alloted}$, the available section of divider array. This is $O(\frac{n}{\log n})$ where $n$ is the size of this recursion instance. Therefore, for the recursive steps, we have the following recursion for work. For problem size $\tilde{n}$ and some positive constant $c_1$:

$$T_1'(\tilde{n}) \leq \tilde{n}^{1/6}T_1'(\tilde{n}^{5/6}) + c_1\left(\frac{\tilde{n}}{\log \tilde{n}}\right)$$

Since we accounted for the base cases earlier, we can set $T_1'(\log n) = 0$, We get that total work in $T_1'(n)$ is at most $\frac{c_1 n \log \log n}{\log n}$ which is clearly dominated by $O(n)$. Therefore, $T_1(n) = T_1'(n) + O(n) + O(n) = O(n)$

For cache misses, we know that the serial algorithm on the $O(\log n)$ sized subproblems incurs $O(n/B)$ cache misses. From Lemma \ref{lma:markingpoints}, we know that marking the points can be amortized to $O(n/B)$ by charging inefficiently marked points to the binary searches. The recursive layers are cache efficient with the exception of $O(n^{2/6} \log n)$ binary searches (and other factors dominated by this). The cache efficient routines incur at most the work divided by $B$, which is $O(n/B)$. The binary searches incur misses until the recursion size is $M$ or $\log n$, whichever is greater. If $\log n$ is greater, then simply bounding by the total work done in the recursions suffices, since for sufficiently large $n$ and any positive constant $c$, $c\frac{n \log \log n}{ \log n} \leq c\frac{n \log M}{M}\leq c\frac{n M^{1/2}}{M} = c\frac{n }{M^{1/2}}$ . By tall cache assumption i.e. $M = \Omega(B^2)$, and thus $\frac{1}{M^{1/2}} = O(\frac{1}{B})$ , the cache misses can be bounded by $O(\frac{cn}{B})=O(n/B)$.

If $M > \log n$, then we consider the following recurrence for subproblems of size $\tilde{n}$:

$$Q(\tilde{n}) = \left\{ \begin{array}{ll}
    O(1 + \frac{\tilde{n}}{B}) & \text{if $\tilde{n} \le M$} \\
     \tilde{n}^{1/6} Q(\tilde{n}^{5/6}) + O(\tilde{n}^{2/6}\log \tilde{n}) & \text{otherwise} 
\end{array}   \right. $$

Consider the following expansion of the recurrence to $k$ terms. For some positive constant $c_2$:
\begin{align*}
    Q(n) &\leq n^{1/6} Q(n^{5/6}) + c_2 n^{2/6}\ln n \\
    & \leq n^{1-(5/6)^k}Q(n^{(5/6)^k}) + c_2 n \ln n \sum_{i=0}^{k-1} \frac{(5/6)^i}{(n^{10/6-1})^{(5/6)^i}}
\end{align*}
Let $R = n^{(5/6)^k}$. We know that when the size of a subproblem is less than or equal to $M$, the subproblem is simply loaded into the cache as the algorithm runs, incurring no more than an additional $O(n/B)$ cache misses across all subproblems of size $\leq M$. We want $R$ to be the size of the largest such subproblem.
\begin{align*}
    R &= n^{(5/6)^k} \leq M < n^{(5/6)^{k-1}}\\
    & \implies M < R^{\frac{6}{5}}\\
    & \implies \frac{1}{R} < \frac{1}{M^{5/6}}
\end{align*}
Let $S = \sum_{i=0}^{k-1} \frac{(5/6)^i}{(n^{10/6-1})^{5/6i}}$ . We know that if the function $f(i) = \frac{(5/6)^i}{(n^{10/6-1})^{5/6i}}$ is increasing from $i$ to $k-1$, then we can use an integral to bound $S$. $f'(i) = n^{-(5/6)^i(10/6-1)} \ln (\frac{6}{5}) \left( (5/6)^{2i} \ln (n^{10/6-1} - (5/6)^i \right)$ is positive for $i \leq \log _{6/5} \ln n - \log _{6/5} \left( \frac{1}{10/6-1} \right)$ . Therefore, in order to bound $S$ by the integral of $f(i)$ from 0 to k, it must be that $k < \log _{6/5} \ln n - \log _{6/5} \left( \frac{1}{10/6-1} \right)$. We know that $R =n^{(5/6)^k}$, which implies $k = \log _{(6/5)} \log _{R} (n)$. It suffices to show that $\log _{(6/5)} \log _{R} (n) <  \log _{6/5} \ln n - \log _{6/5} \left( \frac{1}{10/6-1} \right)$.  
\begin{align*}
   \log _{6/5} \ln n - \log _{6/5} + \log _{6/5} \left( \frac{1}{10/6-1} \right) & < \log _{6/5} \ln n \\
  \implies \ln R &> \frac{1}{2 (\frac{5}{6}) - 1} \\
   \implies R &> e^{1/(10/6 -1)} \approx 4.48169
\end{align*}

Since $f(i)'$ is always increasing over this range, we can use an integral to bound S. $S = \sum_{i=0}^{k-1} \frac{(5/6)^i}{(n^{10/6-1})^{5/6i}} \leq  \int_{0}^{k} \frac{(5/6)^x}{(n^{10/6-1})^{5/6x}}dx$. The integral solves to $\frac{R^{1-10/6}-n^{1-10/6}}{(10/6 - 1) \ln (6/5) \ln n}$. We simplify this.
\begin{align*}
    S \leq \frac{R^{1-10/6}-n^{1-10/6}}{(10/6 - 1) \ln (6/5) \ln n} &< \frac{1}{(10/6-1) \ln (6/5)} \frac{1}{M^{5/6(10/6-1)} \ln n}
    \end{align*}
    Since $5/6 (10/6-1) > 1/2$ and $M = \Omega(B^2)$:
    \begin{align*}
         S < & \frac{1}{(10/6-1) \ln (6/5)} \frac{1}{\Omega(B) \ln n} \\
    & = O \left( \frac{1}{B \ln n} \right)
\end{align*}
Since $S = O(\frac{1}{B \ln n})$, which means the total number of cache misses is $Q(n) = O(n/B) +  c_2 n \ln n \cdot O(\frac{1}{B \ln n}) = O(n/B)$.
\end{proof}

\subsection{Additional Cache Misses from Work-Stealing and False Sharing}\label{sec:cachepresort}

\begin{theorem}
Algorithm \ref{alg:presorted} incurs $O(\frac{n}{B} + S \cdot \frac{M}{B})$ cache misses (including false shares) under a work-stealing scheduler, where $S$ is the number of steals.
\end{theorem}

In the previous discussion of the number of cache misses incurred across the runtime of the algorithm, the only remaining source of cache misses is the scheduler. We use the same notion of work stealing as in Cole and Ramachandran \cite{ColeRamachandran}, ie that in parallel sections, each processor places tasks into a work queue, and where idle processors "steal" work from the tops of these queues to decrease runtime. Associated with work-stealing is an additional type of cache misses, known as false sharing. False sharing occurs in private cache settings when a processor $P_1$ writes to a block which had previously been stored in the cache of another processor $P_2$ and then $P_2$ now has to access that block. Each step in Algorithm \ref{alg:presorted} corresponds to a set of parallel tasks, with synchronization after the conclusion of each step to ensure correctness. Different tasks only share data in the binary searches, where processors may concurrently reading from the same locations, but do not incur any false shares since nothing is updated. However, false shares can result from data alignment. 

For cache-oblivious algorithms, there is no guarantee that subproblems will be partitioned along block boundaries. This can incur at most 1 additional cache miss per boundary due to false sharing. If we implement each step recursively, then we can ensure that these boundaries will only contribute a constant factor of additional false shares before steals. This applies to marking points and setting the WRAPPER flags of the dividers, but is redundant for the merging of dividers as Blelloch and Gibbons \cite{BlellcohGibbons} is already implemented recursively.

Steals are another source of false shares. All of our steps which write to data do so in contiguous array locations within each parallel task. Tasks are ordinarily distributed to processors such that the write locations for successive tasks will be contiguous to each other. However, this contiguity can be disrupted by a steal. Even though a thief may update a piece of memory that won't be needed by the original processor, it may be within a block that is within the original processor's private cache, incurring a false share.  Any particular write can affect at most $B-1$ other processors which are operating in the same area, meaning that false shares only contribute $O(S \cdot B) \subset O(S \cdot \frac{M}{B})$ additional cache misses.

Aside from false shares, the only additional cache misses incurred due to steals are those incurred due the stealing processors accessing data that ordinarily would have been accessed by the original processor in order. This could be as bad as a stealing processor completely reloading as many blocks as could fit into the cache, implying at worst $O(S \cdot \frac{M}{B})$ cache misses. So in total, when using any work-stealing scheduler, there is at most $O(\frac{n}{B} + S \cdot \frac{M}{B})$ cache misses incurred by Algorithm \ref{alg:presorted}.

\begin{corollary}
Algorithm \ref{alg:presorted} incurs $O(n/B)$ cache misses \textit{w.h.p} in $p$ with the RWS scheduler when the number of processors is $p = O(\frac{n}{M \log n})$. 
\end{corollary}

We additionally consider a randomized work-stealing scheduler, the same as is described in Cole and Ramachandran \cite{ColeRamachandran} and which was originally described by Blumofe and Leiserson \cite{BlumodeLeiserson}. The randomized work-stealing scheduler (RWS) is a work-stealing scheduler where "idle" processors steal the head task from a randomly selected processor. If this fails, as in the task queue selected is empty, the scheduler simply tries again until either a task is found or the parallel section is completed.

Acar, Blelloch, and Blumofe \cite{acar2000data} show that w.h.p. in $p$, the number of steals incurred during execution using the RWS is $S = O(p \cdot T_\infty)$. Therefore, if $p = O(\frac{n}{M \log n})$, then $S = O(\frac{n}{M})$, which when applied to the total number of cache misses yields $O(\frac{n}{B} + \frac{n}{M} \cdot \frac{M}{B}) = O(n/B)$.

%% file: algorithm_docs/multiwayMerge.tex
\section{A Cache- and Work-Optimal \texorpdfstring{$\Theta(\log{n}\log{\log{n}})$}{Theta(log n log log n)} Span Algorithm}\label{sec:multiway}


The current state of the art for deterministic parallel sorting algorithms in the binary-forking model with regards to span is Cole and Ramachandran's \cite{ColeRamachandran} $O(\log n \log \log n)$ span cache oblivious algorithm. Since the algorithm is cache optimal, costing only $O((n/B) \log _M (n))$ cache misses, this means that using their algorithm to sort an instance of unsorted points, followed by our Algorithm \ref{alg:presorted} (pseudocode in appendix)
 will yield a work and cache-optimal algorithm with $O(\log n \log \log n)$ span.
 
However, considering that implementation of Cole and Ramachandran's sorting algorithm is already required for this, a simpler approach for unsorted points is to modify the internal structure of the sorting algorithm. Overall, this modification is much simpler than implementing the presorted algorithm after already implementing Cole and Ramachandran's. We propose a divide-and-conquer algorithm for unsorted points which performs $O(n \log n)$ work in $O(\log n \log \log n)$ span while incurring only $O((n/B)\log_M n)$ cache misses. Our modifications are to account for the fact that sorting alone does not remove points which are not extreme points of the convex hull. Our augmentations preserve many key properties of Cole and Ramachandran's algorithm. Notably, ours is also cache-oblivious and cache optimal (From Arge and Miltersen \cite{ArgeMiltersen}, when output sensitivity is not a consideration). The cache complexity of our algorithm matches Sharma and Sen's \cite{SharmaSen} randomized cache oblivious convex hull algorithm but does so in the worst case rather than in expectation.

Cole and Ramachandran's \cite{ColeRamachandran} cache-oblivious sorting algorithm is a divide-and-conquer algorithm with a relatively complicated recursive structure. It takes as input $r$ sorted lists of elements with total size $n$. Depending on the relative size between $n$ and $r$, the algorithm reorganizes the data so that elements of similar rank are grouped together into subproblems. This is accomplished by using a set of pivots which is of size approximately $n/r^3$ if $r$ is sufficiently large. These subproblems are divided further into smaller subproblems which are then merged together. At this point the entire input has been sorted because the initial pivot-based partition ensures that for any two (larger) subproblems the ranks of all elements in one are larger than those of all elements in the other.

Our algorithm determines what we refer to as the "quarter-hull", and is thus iterated 4 times to produce the complete convex hull. We define the "quarter-hull" of a point set $S$ as the counter-clockwise section of the convex hull between the point in $S$ with the largest y-coordinate and the point in $S$ with smallest x-coordinate. Since the convex hull is preserved under dimension-preserving linear transformations, it is clear that this algorithm can be applied to all 4 "quadrants" of the convex hull by rotation. It can also be the case that the smallest x-coordinate and largest y-coordinate points are the same point, in which case the quarter hull is only that singleton point and can be returned immediately. In case of a tie for x-coordinate, use the largest y-coordinate and vice versa.

The data structure is much simpler than in the presorted algorithm. For each point, $p$, we simply need to include an additional $(x,y)$ coordinate pair, which we will refer to as the TAIL, and two integers RANK, and HOME. If the TAIL lies to the left and below $p$, this indicates that it is an extreme point, otherwise it is not. Implicitly, the TAIL defines an "active range" for $p$ starting with it predecessor in the hull. However, the TAIL is not a pointer, it explicitly stores the coordinates of the predecessor in the hull. We will refer to $p$ itself as the HEAD of the point. RANK will be used to perform all prefix sums and HOME will mark the "true" subset that it would be placed in if the TAIL were the same as the point itself.

\begin{algorithm}[ht]
\small
\caption{Divide-and-Conquer with Multiway Merge}\label{alg:MultiwayMerge}
\KwData{$A$, a collection of quarter hulls, $L_1, L_2, ..., L_r$ where $n\leq 3r^6$}
\KwResult{The combined quarter hull of $A$}
\textbf{if} $n\leq 24$ \textbf{then} apply any serial convex hull algorithm and return\;
\textbf{if} $n\leq 3r^3$ \textbf{then} $k \gets 1, A_1 \gets A$
\Else {
Form a sample $S$ of every $r^2$th point in each $L_i$, for a total of $n_i/r^2$ points from each quarter hull, where $n_i$ is the number of extreme points defining quarter hull $L_i$\;
Compute ranks of elements of $S$ using TAIL\;
Form a sample $P$ of every $2r$th member in $S$ by rank for a total of $\leq n/2r^3$ elements\;
Using $P$ as a set of x-coordinate pivots (by TAIL), partition $A$ into $k=|P|+1$ subsets ($A_1, ... A_k$). Include the points for which the x-coordinate interval between its TAIL and itself contains some overlap with the interval between pivots $P[i-1],P[i]$ (See Figure 6)\;
}
\textbf{parallel }\ForEach {subset $A_i$ of $A$ (from $1$ to $k$) }
{
    Separate $A_i$ into smaller subsets $A_{ij}$ such that each contains elements from at most $\sqrt{r}$ different lists\;
    \textbf{parallel for each }$A_{ij}$ \textbf{do} MulitwayMerge($A_{ij}$) \; Run MulitwayMerge($A_i$) using the sorted $A_{ij}$ as lists\;
}   
\textbf{parallel} \ForEach{$p \in [A_1, A_2, ... A_k]$}{
    Mark any point for which its active range no longer overlaps with the subset it is assigned to.
}
Using prefix sums on each subset, eliminate any marked points.
\textbf{parallel} \ForEach{$i \in [1,k]$}{
     Determine if the first element of $A_{i+1}$ invalidates the last element of $A_{i}$. If so, set the TAIL of the first element of $A_{i+1}$ to the second to last element of $A_i$. Otherwise, set its TAIL to the last element of $A_i$.
}
Using prefix sums, reassign all the identified extreme points into adjacent sorted positions in an output array.
\end{algorithm}

\begin{theorem}
For $n$ points in the plane, Algorithm~\ref{alg:MultiwayMerge} (Multiway Merge) finds their quarter hull in $O(n \log n)$ work and $O(\log n \log \log n)$ span.
\end{theorem}
\begin{proof}
    
Our algorithm differs from Cole and Ramachandran's \cite{ColeRamachandran} in two key ways. The first is the size of the subsets $A_i$, since ours includes points which may be on the other side of the pivot but their TAILs are on the correct side, there can be at most $r$ more elements in each of our subsets than in the original sorting algorithm. This is because each list can only contribute 1 copy of a point which "belongs" in another subset since the lists are given as quarter hulls. However, since the size of their subsets is at most $3r^3-r^2-r$ (From Lemma 2.1 of \cite{ColeRamachandran}) and the bound required for the recursion to hold is that the subsets be of size at most $3r^3$, this will not be an issue. Ours will still work since $3r^3-r^2\leq 3r^3$. This overlap also presents marginal difficulty in the implementation of the partition step. Within each of the $r$ lists, it is generally the case that a particular pivot will lie between the HEAD and TAIL of a point, leading to that point being represented in more than one subset in the partition. To achieve this, Cole and Ramachandran find the element of each list which is just after the point in the ordering to determine which subsection of values needs to be copied into each bucket. We can simply buffer these ranges for each bucket and we will be sure to capture all the required points, incurring specifically the extra $r$ for each bucket.

The other difference is that we have to check the boundaries of each subset to make sure that the last point in some quarter hull $A_i$ is not invalidated by $A_{i+1}$ and vice versa. That is determined by observing the quarterhull of the boundary points and their TAILs (see Figure \ref{fig:boundarycheck}). Update the TAILs of the points to be equal to the predecessor along this quarter-hull. If a point is deleted, update it such that the new first point in $A_{i+1}$ has the new last point in $A_{i}$ as its TAIL.

\begin{figure}[t]
    \centering
    \begin{subfigure}[b]{0.6\textwidth}
        \centering
        \includegraphics[height=0.2\textwidth]{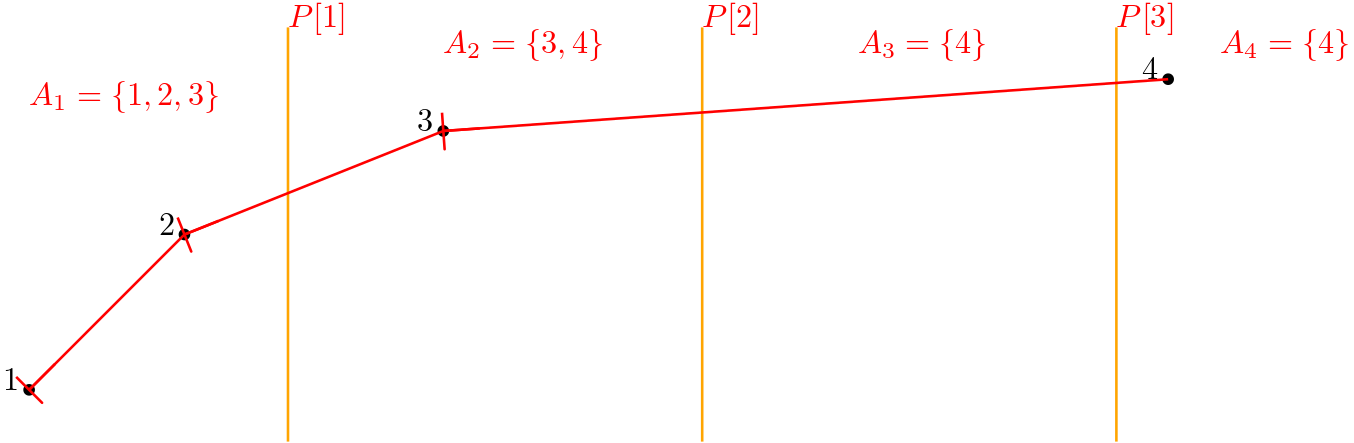}
     \caption{}
     \label{fig:pivotc}
    \end{subfigure}
    \hfill
    \begin{subfigure}[b]{0.35\textwidth}
        \centering
        \includegraphics[width=.7\textwidth]{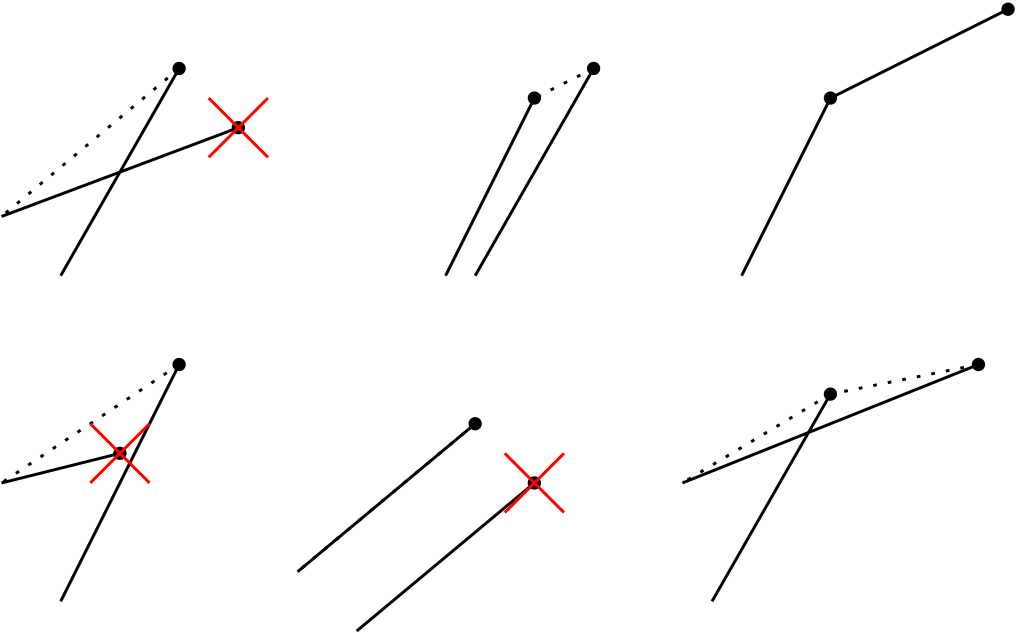}
        \caption{}
        \label{fig:boundarycheck}
    \end{subfigure}
    \caption{(a) Showing how the pivots work in Algorithm~\ref{alg:MultiwayMerge}; (b) Possible configurations of points on the boundaries between subsets in Algorithm ~\ref{alg:MultiwayMerge}. TAILs before the check are shown with solid lines and updated TAILs are with dotted lines.}
    \label{fig:alg2}
\end{figure}

We do not need to check any more than the boundaries. Since we delete from the subsets points whose intervals are outside the subset's interval induced by the pivots, we know that the only points that can have intervals which are partially outside the subset's interval are at the boundaries. We also know that every point's interval is disjoint from those of the other points except for the x-coordinate at which one point's TAIL coincides with the HEAD of another. 

This check takes $O(\log r) \le O(\log n)$ span and performs $O(n/r^3) \le O(\sqrt{n})$ work, which is dominated by the work and span of the rest of the algorithm. Therefore, the work and span bounds of our algorithm are equivalent to Cole and Ramachandran's sorting algorithm \cite{ColeRamachandran}, $O(n \log n)$ and $O(\log n \log \log n)$, respectively.\end{proof}

\begin{theorem}
For n points in 2D, Algorithm \ref{alg:MultiwayMerge} finds their convex hull while incurring at most $O((n/B) \log _M n)$ cache misses.
\end{theorem}

\begin{proof}
    This follows directly from the Cole and Ramachandran bounds. We use the same tall cache assumption as before, that $M = \Omega (B^2)$. At each recursion layer, at most $O(n/B)$ cache misses are incurred. This is trivially true for the elimination of points who are not active within their subset's ranges (steps 12 and 13 in Algorithm \ref{alg:MultiwayMerge}). It is also trivial for the elimination phase at the end of the execution (step 15).
    
    The additional copies of points which belong in multiple buckets also incur at most $O(n/B)$ additional cache misses. The reads required to make these additional copies are contiguous with the other points which are being placed into that bucket, so it can be at worst an additional $O(n/B)$ cache misses, since it would have the exact same misalignment issues as Cole and Ramachandran's.
    
    For the comparisons in Step 14, we access at most $s / r$ memory locations, where $s$ is the size of the sample S.  Since $s / r < s \cdot r$, then even if every memory access in the Step 14 is a cache miss, we will not exceed $O(n/B)$ cache misses at each recursion layer. Hence, the recursion for Cole and Ramachandran's sorting algorithm also applies here: $Q(n,r) = O(n/B) + \sum _{i,j}Q(n_{ij}, \sqrt{r}) + \sum Q(n_i, \sqrt{r})$ which solves to $Q(n,r) = O(\frac{n \log n}{B \log M})$
\end{proof}

\begin{theorem} \label{theorem:workstealmulti}
Using any work-stealing scheduler, Algorithm \ref{alg:MultiwayMerge} incurs at most $O((n/B) \log _M n + S\cdot \frac{M}{B})$ cache misses where $S$ is the number of steals.
\end{theorem}
\begin{proof}

This is the exact same bound as is given in Cole and Ramachandran \cite{ColeRamachandran}, so it suffices to show that at most $O(M/B)$ cache misses are incurred when a processor steals one of the modified tasks. The sampling and partition phase (Steps 2-6 in Algorithm \ref{alg:MultiwayMerge}) is very similar to Cole and Ramachandran's, and incurs the same number of cache misses due to steals and false sharing asymptotically. This follows from the fact that each parallel task from the modified procedure in Step 6 only makes a constant number of writes, incurring at most $O(S \cdot B) \subset O(S \cdot \frac{M}{B})$ (due to the tall cache assumption) false shares. 

The decomposition into even smaller subproblems (Steps 7-10) uses the same structure as Cole and Ramachandran's sorting algorithm and thus is already know to incur an additional $O(S\cdot \frac{M}{B})$ cache misses. All that remains is the elimination of points which extend outside the bucket they were placed in during partitioning (Steps 12-13), checking the boundaries (Step 14), and the final deletion (Step 15). Both rounds of deletions also perform only a constant number of writes, which means that only $O(S\cdot \frac{M}{B})$ false shares can be incurred. The exact same bound applies to the boundary checks.  

Aside from the cache misses incurred by false shares, the cache misses incurred from performing a steal can be at most $M/B$, since any more cache misses would be incurred without steals. This means that in total, there will be $O(S\cdot \frac{M}{B})$ additional cache misses due to the use of a work-stealing scheduler.\end{proof}

\begin{corollary}
Algorithm \ref{alg:presorted} incurs $O(n/B \log_M (n))$ cache misses \textit{w.h.p} in $n$ with the RWS scheduler when the number of processors is $p = O  \left(\frac{n}{M \log M} \cdot \text{min } \left( \frac{1}{\log \log n}, \frac{M \log B}{B^2} \right)  \right)$. 
\end{corollary}\begin{proof}

This comes directly from our Theorem \ref{theorem:workstealmulti}
and Theorem 1.2 \cite{ColeRamachandran}. Since our algorithm matches the false sharing and cache miss bounds of Cole and Ramachandran's, their analysis directly carries over. \end{proof}

%% file: conclusion.tex
\section{Conclusion}\label{sec:conclusion}
We have presented two cache-oblivious algorithms for the binary-forking model which find the convex hull of a set of points in 2 dimensions. We presented the first cache-oblivious algorithm for presorted points which achieves optimal span $O(\log n)$ with optimal work, $O(n)$,  while incurring the optimal amount of cache misses $O(n/B)$. The current state of the art for sorting in the binary forking model achieves $O(\log n \log \log n)$ span with $O(n \log n)$ work and while incurring $O(n/B \log _M(n))$ cache misses. Here only span is suboptimal when considering only the size of the input (as opposed to also considering the size of the output). By modifying this algorithm, we provide a simpler method to achieve these bounds for convex hulls compared to sorting and then applying any algorithm for presorted points that we are aware of. We believe our techniques for converting merge-based sorting algorithms into convex hull algorithms can be generalized to many merge-based sorting algorithms. Further work would entail expanding the algorithm for presorted input to the more general case of simple polygons, reducing the span of sorting in the binary forking model and  developing a cache-oblivious algorithm which achieves work which is optimal with respect to the size of the output as well as the input. 